\documentclass[sigconf,]{acmart}
\usepackage{multirow}
\usepackage{graphicx}
\usepackage{xspace}
\usepackage{amsthm}
\usepackage{amsmath}
\usepackage{enumitem}
\usepackage{booktabs}
\usepackage{bbding}
\usepackage{float}
\usepackage{caption}
\usepackage{subcaption}
\usepackage{balance}
\usepackage{breakurl}
\usepackage{algorithm}
\usepackage{algorithmicx}
\usepackage{algpseudocode}

\usepackage{bbm}

\AtBeginDocument{%
  \providecommand\BibTeX{{%
    \normalfont B\kern-0.5em{\scshape i\kern-0.25em b}\kern-0.8em\TeX}}}

\setcopyright{acmcopyright}
\copyrightyear{2018}
\acmYear{2018}
\acmDOI{XXXXXXX.XXXXXXX}

\acmPrice{15.00}
\acmISBN{978-1-4503-XXXX-X/18/06}

\begin{document}
\title{Confidence-aware Fine-tuning of Sequential Recommendation
Systems via Conformal Prediction}

\author{Chen Wang}
\email{cwang266@uic.edu}

\orcid{0000-0001-5264-3305}
\affiliation{%
  \institution{University of Illinois Chicago}
  \country{USA}
}

\author{Fangxin Wang}
\email{fwang51@uic.edu}
\affiliation{%
  \institution{University of Illinois Chicago}
  \country{USA}
}

\author[]{Ruocheng Guo}
\authornote{Work not related to ByteDance.}
\email{rguo.asu@gmail.com}
\affiliation{%
  \institution{ByteDance Research}
  \country{UK}
}



\author[]{Yueqing Liang}
\email{yliang40@hawk.iit.edu}
\affiliation{
  \institution{Illinois Institute of Technology}
  \country{USA}
}

\author[]{Philip S. Yu}
\affiliation{%
  \institution{University of Illinois Chicago}
  \country{USA}}
\email{psyu@uic.edu}

\renewcommand{\shortauthors}{Trovato and Tobin, et al.}

\begin{abstract}
In Sequential Recommendation Systems (SRecsys), traditional training approaches that rely on Cross-Entropy (CE) loss often prioritize accuracy but fail to align well with user satisfaction metrics. CE loss focuses on maximizing the confidence of the ground truth item, which is challenging to achieve universally across all users and sessions. It also overlooks the practical acceptability of ranking the ground truth item within the top-$K$ positions, a common metric in SRecsys. To address this limitation, we propose \textbf{CPFT}, a novel fine-tuning framework that integrates Conformal Prediction (CP)-based losses with CE loss to optimize accuracy alongside confidence that better aligns with widely used top-$K$ metrics. CPFT embeds CP principles into the training loop using differentiable proxy losses and computationally efficient calibration strategies, enabling the generation of high-confidence prediction sets. These sets focus on items with high relevance while maintaining robust coverage guarantees. Extensive experiments on five real-world datasets and four distinct sequential models demonstrate that CPFT improves precision metrics and confidence calibration. Our results highlight the importance of confidence-aware fine-tuning in delivering accurate, trustworthy recommendations that enhance user satisfaction.
\end{abstract}





\maketitle
\section{Introduction}
\label{sec:intro}

\begin{figure}[t]
    \centering
    \includegraphics[width=\linewidth]{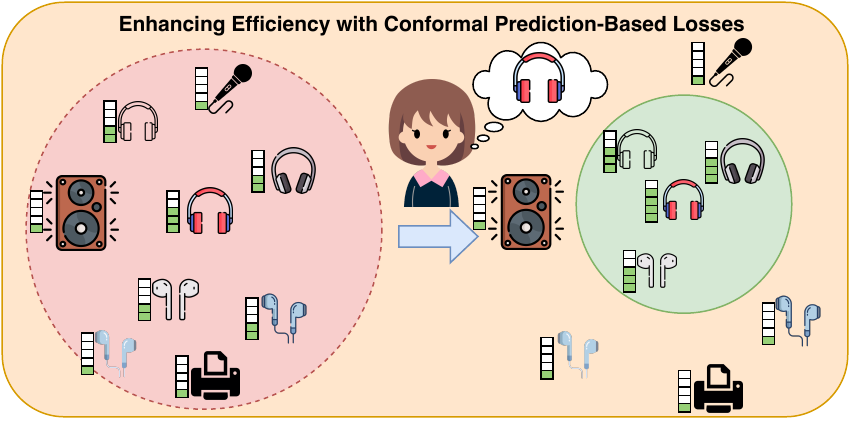}
    \caption{Conceptual overview of CPFT. By combining \emph{confidence} (via CP-based losses) with standard cross-entropy objectives, CPFT produces a smaller, high-confidence prediction set. The green shading reflects item-level confidence scores, which become more concentrated on relevant items (circles) under CP-based training.}
    \label{fig:intro}
\end{figure}

Personalized recommendation systems are fundamental to enhancing user experiences across e-commerce, streaming services, and social platforms. Sequential Recommendation Systems (SRecsys)~\cite{sasrec,s3rec,fdsa,UnisRec} have demonstrated strong capabilities in leveraging user--item interaction histories to predict a user's next item of interest. These systems improve user engagement, satisfaction, and business revenue by providing accurate and timely recommendations. However, while advances in sequential modeling have significantly enhanced prediction accuracy, the critical aspect of \emph{confidence} in recommendations remains underexplored.

Traditional approaches primarily rely on Cross-Entropy (CE) loss~\cite{sasrec, s3rec, sun2019bert4rec} to optimize for metrics such as hit rate and recall. Although these metrics ensure that recommended items are often accurate, they do not account for the confidence level associated with each prediction. This distinction is crucial, as user satisfaction depends not only on receiving the correct recommendation but also on trusting the system's confidence in its predictions~\cite{user-saticfaction}.

Confidence in this context reflects the system's certainty in its recommendation. For example, a recommendation with 95\% confidence indicates the system strongly believes in the item's relevance to the user. High-confidence recommendations foster trust and engagement, while low-confidence predictions may undermine the system's credibility, even when they are correct. The ability to balance accuracy and confidence is thus essential for improving both the reliability and user satisfaction of recommendation systems.

To address this issue, we draw upon Conformal Prediction (CP)~\cite{vovk2005algorithmic, angelopoulos2021gentle}, a statistical framework that generates \emph{prediction sets}—lists of items guaranteed, under certain assumptions, to contain the correct item with high probability. In recommendation systems, the size of these prediction sets encodes the model's uncertainty: smaller sets reflect higher confidence, while larger sets indicate ambiguity. For instance, CP might recommend a set containing a red headphone and a blue headphone with a 90\% confidence guarantee, balancing precision and reliability. CP has shown success in tasks like classification~\cite{angelopoulos2020uncertainty} and graph modeling~\cite{huang2023graph}, but its use in recommendation systems has been limited to inference-time applications~\cite{angelopoulos2023recommendation}. Integrating CP into the training process is challenging due to its non-differentiability and the computational cost of generating prediction sets for large-scale datasets.

To overcome these challenges, we propose \textbf{Confidence-aware Fine-Tuning (CPFT)}, a novel fine-tuning strategy for SRecsys models that incorporates CP principles into the training loop. CPFT introduces differentiable \emph{proxy losses} to approximate CP objectives, enabling backpropagation during training. Additionally, a calibration-based approximation reduces computational overhead by leveraging subsets of validation sequences. As shown in Figure~\ref{fig:intro}, CPFT refines confidence distributions, producing smaller and more accurate prediction sets by emphasizing high-relevance items.

Notably, our CPFT loss serves not only as a training objective but also as a confidence metric. From a CP perspective, the Conformal Prediction Set Size (CPS) loss captures the confidence associated with a recommendation by minimizing the size of the prediction set, with smaller sets reflecting greater confidence. Similarly, the Conformal Prediction Set Distance (CPD) loss ensures that top-ranked items in the set are closely aligned with the ground truth, reinforcing confidence in the model’s most relevant predictions. Together, CPS and CPD provide a unified signal that reflects both the precision and reliability of the model’s predictions, enabling CPFT to balance top-1 accuracy with confidence allocation over good items.

Our approach introduces two CP-inspired losses: \textbf{Conformal Prediction Set Size (CPS)} and \textbf{Conformal Prediction Set Distance (CPD)}. CPS penalizes large prediction sets, encouraging selective recommendations and reducing cognitive load on users. CPD minimizes the risk of excluding correct items by drawing top-ranked candidates closer to the ground truth in embedding space. By integrating these losses as regularizers alongside CE loss, CPFT enhances confidence while building on pre-trained models. Furthermore, CPFT utilizes validation data for conformal calibration, injecting additional information about uncertainty without requiring new labels.

In summary, the key contributions of our work are:
\begin{enumerate}[leftmargin=*]
    \item We introduce the integration of \emph{conformal prediction-based losses} into Sequential Recommendation Systems (SRecsys), marking a novel approach for balancing top-1 accuracy and confidence allocation over good items.
    \item We propose two complementary CP-based losses—\textbf{CPS} and \textbf{CPD}—that can be easily \emph{plugged} into any Transformer- or RNN-based sequential model trained via cross-entropy.
    \item We demonstrate the effectiveness of our method across five real-world datasets and four distinct types of SRecsys, showcasing improvements in both precision metrics and confidence calibration.
    \item We overcome key barriers to integrating CP into training by introducing differentiable approximations and computationally efficient calibration strategies.
\end{enumerate}

\section{Related Work}

\subsection{Sequential Recommendation}
Sequential Recommendation Systems (SRecsys) predict users’ evolving preferences based on their historical interaction sequences. Early methods relied on Markov Chains~\cite{rendle2010factorizing, he2016fusing} and RNNs~\cite{hidasi2015session,ma2019hierarchical} to capture temporal dynamics, but Transformers have emerged as the dominant approach due to their ability to model complex item transitions and long-range dependencies. SASRec~\cite{sasrec} demonstrated the effectiveness of Transformers for SRecsys, while BERT4Rec~\cite{sun2019bert4rec} introduced bidirectional attention to improve contextual modeling. Recent methods, such as CoSeRec~\cite{liu2021contrastive} and ASReP~\cite{liu2021augmenting}, leverage self-supervised learning (SSL) to enhance sequence representations.

Despite these advancements in accuracy, most existing methods neglect confidence—a crucial factor for building user trust. Our approach addresses this gap by incorporating Conformal Prediction (CP)-based losses, which balance recommendation accuracy with confidence, ultimately improving both predictive performance and user satisfaction.

\subsection{Confidence in Recommendation Systems}
Confidence in recommender systems (RecSys) pertains to how certain a model is about the relevance of recommended items. Incorporating confidence metrics can bolster user trust and satisfaction, as it ensures the system not only predicts items users are likely to engage with but also communicates its certainty. For instance, McNee et al.~\cite{mcnee2003confidence} showed that displaying a confidence metric positively influences user behavior, underscoring the importance of transparent recommendation. Another example is the confidence-aware re-ranking algorithm proposed by Naghiaei et al.~\cite{naghiaei2022towards}, which jointly optimizes calibration, relevance, and diversity by factoring in user profile size as a proxy for confidence. Likewise, Gohari et al.~\cite{gohari2018new} introduced a Confidence-Based Recommendation (CBR) framework leveraging trust and certainty for collaborative filtering, while Azadjalal et al.~\cite{azadjalal2017trust} highlighted that confidence estimates can mitigate data sparsity by emphasizing implicit trust signals. 

Existing studies typically focus on heuristics or re-ranking strategies for confidence calibration, which are applied post-hoc and lack the systematic optimization of confidence during training. In contrast, our work introduces \emph{CP-based losses} that systematically optimize the overall confidence within the model’s top-ranked recommendations. By minimizing the size of the prediction set (and/or its distance to the ground truth), our approach achieves high coverage with smaller sets, thus offering robust, confidence-aware recommendations during both training and inference. To address these gaps, we leverage Conformal Prediction, which we detail below.

\subsection{Conformal Prediction}
\label{Background: Conformal Prediction}
Conformal Prediction (CP)~\cite{gammerman1998learning,vovk2005algorithmic} is a model-agnostic framework that provides distribution-free prediction set guarantees. Split CP~\cite{papadopoulos2002inductive}, a common variant, uses a hold-out validation set to calibrate the model and ensure statistical coverage. CP has been widely applied in domains such as image classification~\cite{angelopoulos2020uncertainty}, drug discovery~\cite{alvarsson2021predicting}, and question answering~\cite{fisch2020efficient}. Recent research has also explored integrating CP into training to optimize inefficiency (i.e., prediction set size)~\cite{stutz2021learning,huang2023graph}, though this often involves a trade-off between coverage and accuracy.

In recommender systems, CP has predominantly been applied at inference or through specialized model designs. For example, Angelopoulos et al.~\cite{angelopoulos2023recommendation} proposed a CP-based learning-to-rank framework to enhance diversity and control false discovery rates, while Kagita et al.~\cite{KAGITA2022109108} developed an Inductive Conformal Recommender System (ICRS) to improve confidence calibration using nonconformity measures derived during training. Similarly, a Hybrid Content-based Fuzzy Conformal Recommender System (HCF-CRS)~\cite{10.1371/journal.pone.0204849} addressed data sparsity by combining fuzzy logic with CP to assign confidence scores to recommendations.

Unlike prior approaches, which primarily apply CP at inference or rely on bespoke designs, our work \emph{integrates CP objectives as trainable losses} within \emph{Sequential Recommendation Systems (SRecsys)}. We introduce two novel CP-based losses—\emph{Conformal Prediction Set Size (CPS)} and \emph{Conformal Prediction Set Distance (CPD)}—and incorporate them alongside cross-entropy (CE) loss during fine-tuning. This enables the model to generate compact, high-confidence recommendation sets that improve user satisfaction without sacrificing predictive accuracy.

\section{Preliminaries}
\label{sec:Preliminaries}
In this section, we outline the standard setup of Sequential Recommendation Systems (SRecsys), review the commonly used Cross-Entropy (CE) loss function, and introduce the concept of Conformal Prediction (CP), which underpins our confidence-aware approach.

\subsection{Sequential Recommendation Systems}
\label{subsec:sequential_rec}
Let $\mathcal{U}$ be the set of users, with each user denoted by $u_i \in \mathcal{U}$, and let $\mathcal{V}$ be the set of items, with each item denoted by $v_j \in \mathcal{V}$. In Sequential Recommendation Systems, each user $u_i$ has a chronological sequence of item interactions
\[
\mathcal{S}^{(i)} = \big[v^{(i)}_1,\, v^{(i)}_2,\, \ldots,\, v^{(i)}_{T_i}\big],
\]
where $v^{(i)}_t \in \mathcal{V}$ is the item that user $u_i$ interacts with at time step $t$, and $T_i = |\mathcal{S}^{(i)}|$ is the length of the sequence for user $u_i$. The central task is to predict the next item $v^{(i)}_{T_i+1}$ for user $u_i$ based on their previous interactions $\mathcal{S}^{(i)}$. Formally, the goal is to estimate
\[
\mathbb{P}\bigl( v_j = v^{(i)}_{T_i+1} \;\big|\; \mathcal{S}^{(i)} \bigr)
\quad \text{for all } v_j \in \mathcal{V}.
\]

Each item $v_j$ is mapped to a $d$-dimensional embedding vector $\mathbf{e}_j \in \mathbb{R}^d$ via an embedding matrix $\mathbf{E} \in \mathbb{R}^{|\mathcal{V}| \times d}$. During inference, a recommendation model $\hat{f}$ encodes a user's interaction sequence into a $d$-dimensional representation:
\[
\mathbf{H}^{(i)} = \hat{f}\bigl(\mathcal{S}^{(i)}\bigr) \;\in\; \mathbb{R}^{1 \times d}.
\]
The relevance score between $\mathbf{H}^{(i)}$ and an item embedding $\mathbf{e}_j$ is then computed as
\begin{equation}
\label{Eq: relevance}
    r(u_i, v_j) = \mathbf{H}^{(i)} \,\cdot\, \mathbf{e}_j^\top.
\end{equation}
These relevance scores are used to rank items so that the system can recommend the most promising candidates for user $u_i$ at the next time step $T_i+1$. Understanding these sequential patterns is critical for generating high-confidence predictions, as they help identify items likely to align with user preferences.

\subsection{Cross-Entropy (CE) Loss}
\label{subsec:CE_loss}
The Cross-Entropy (CE) loss is widely used in SRecsys to measure the discrepancy between the predicted probability distribution over items and the actual next item in a user’s sequence. For user $u_i$ and time step $t$ (where $1 \leq t \leq T_i$), let $v^{(i)}_{t}$ be the ground truth next item. The CE loss is defined as:
\begin{equation}
\mathcal{L}_{\text{CE}}^{(i,t)} 
= - \sum_{v_j \in \mathcal{V}} \mathbbm{1}\bigl[v_j = v^{(i)}_{t}\bigr] 
\log \mathbb{P}\bigl(v_j = v^{(i)}_{t} \mid \mathcal{S}^{(i)}\bigr),
\end{equation}
where $\mathbbm{1}\bigl[v_j = v^{(i)}_{t}\bigr]$ is an indicator function (1 if $v_j$ is indeed the next item $v^{(i)}_{t}$, and 0 otherwise), and $\mathbb{P}\bigl(v_j = v^{(i)}_{t} \mid \mathcal{S}^{(i)}\bigr)$ is the predicted probability that $v_j$ is the next item. While CE loss ensures accurate predictions, it does not account for the model's confidence in these predictions, which is crucial for enhancing user trust and satisfaction.

\subsection{Conformal Prediction}
\label{prelim:cp}
Conformal Prediction (CP)~\cite{gammerman1998learning,vovk2005algorithmic} is a framework for producing prediction \emph{sets} with theoretical coverage guarantees. In the context of sequential recommendation, CP can provide a set of candidate next items such that, with high probability, this set contains the user’s true next item.

\paragraph{Split Conformal Prediction.}
Split CP~\cite{papadopoulos2002inductive} is a common CP variant that partitions data into two subsets: a \emph{training set} for fitting the model $\hat{f}$ and a \emph{calibration set} for estimating nonconformity scores. Nonconformity scores capture how “unlikely” a candidate item $v_j$ is given the model’s prediction $\mathbf{H}^{(i)}$ for user $u_i$. A larger score indicates lower conformity between the item and the model’s predictions.

Once nonconformity scores are obtained on the calibration set, we determine the $(1-\alpha)$-quantile $\hat{q}_{1-\alpha}$, which serves as a threshold. For a user $u_i$ with sequence $\mathcal{S}^{(i)}$, the prediction set is then constructed as
\begin{equation}
\label{Eq: construct}
\hat{C}_{\alpha}\bigl(\mathcal{S}^{(i)}\bigr)
= \Bigl\{
v_j \,\in\, \mathcal{V} : 
s\bigl(\mathbf{H}^{(i)}, v_j\bigr) \;\le\; \hat{q}_{1-\alpha}
\Bigr\},
\end{equation}
where $s\bigl(\mathbf{H}^{(i)}, v_j\bigr)$ denotes the nonconformity score between the representation $\mathbf{H}^{(i)}$ and the candidate item embedding $\mathbf{e}_j$.

\paragraph{Coverage Guarantee.}
Under the assumption of exchangeability between the calibration data and future test instances, CP guarantees that the true next item $v^{(i)}_{T_i+1}$ lies in the set $\hat{C}_{\alpha}\bigl(\mathcal{S}^{(i)}\bigr)$ with probability at least $1 - \alpha$. Formally,
\[
\mathbb{P}\Bigl(v^{(i)}_{T_i+1} \in \hat{C}_{\alpha}\bigl(\mathcal{S}^{(i)}\bigr)\Bigr) 
\;\ge\; 1 - \alpha.
\]
In SRecsys, this translates to a high-confidence set of recommended items, ensuring that the model’s uncertainty about the exact next item is explicitly quantified. Consequently, CP-based approaches enhance the robustness of recommendations by controlling how often the true item falls outside the generated set.

\section{Methodology}
\label{sec:methodology}
\begin{figure*}[ht]
    \centering
    \includegraphics[width=\linewidth]{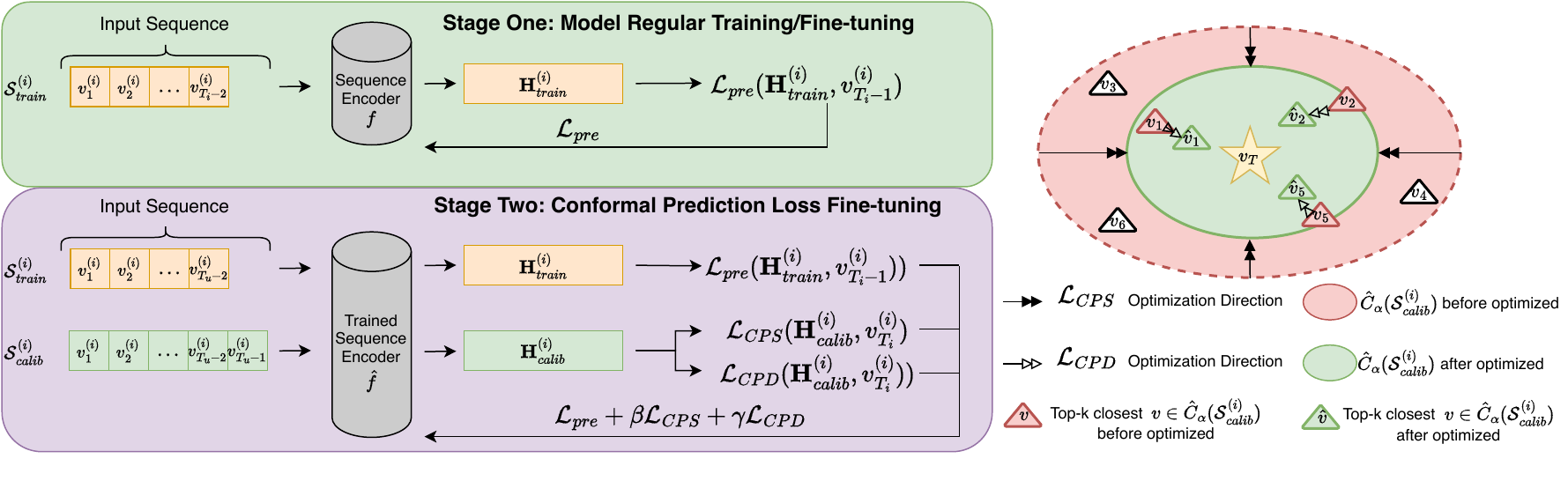}
    \caption{Visualization of the proposed framework. 
    \textbf{Stage One:} Standard training/fine-tuning for a sequential recommendation model. 
    \textbf{Stage Two:} Data splitting partitions each user's sequence into \emph{training} and \emph{calibration} parts. Standard supervised loss (e.g., cross-entropy) is computed on the training sequences, while our CP-based losses (CPS and CPD) are computed on the calibration sequences. The inset (right) illustrates how $\mathcal{L}_{\text{CPS}}$ seeks to reduce the prediction set size, whereas $\mathcal{L}_{\text{CPD}}$ forces the top-$K$ items within the set to be closer to the ground truth item.}
    \label{fig:architec}
\end{figure*}

We present a novel fine-tuning framework for an existing sequential recommendation model, denoted $\hat{f}$, by incorporating the principles of Conformal Prediction (CP). Our objective is to enhance both \emph{accuracy} and \emph{confidence} in the model’s predictions by refining them during training—rather than solely at inference time. In what follows, we introduce our data-splitting procedure, explain how CP can be adapted for sequential recommendation, and detail our new CP-based loss functions (CPS and CPD), which we use to augment standard supervised losses. A schematic illustration of our framework is shown in Figure~\ref{fig:architec}.

\subsection{Motivation and High-Level Idea}
Conformal Prediction (CP) originated as a statistical framework for constructing \emph{prediction sets} that, with high probability, contain the correct outcome. In the context of \emph{recommendation systems}, these prediction sets translate into \emph{ranked lists of items} that should include the true next item in a user’s sequence at least $(1 - \alpha)$ fraction of the time. Unlike common ranking methods that focus only on predicting a single item or that provide no formal notion of ``confidence,'' CP ensures a statistically valid \emph{coverage guarantee}. 

However, prior works have primarily applied CP at \emph{inference}, where it is used to generate confidence-controlled prediction sets without influencing the model’s learning process. Our key insight is that by \emph{integrating CP principles during training}, we can refine the model’s parameters to inherently produce smaller, more confident prediction sets. This integration reduces uncertainty in recommendations while preserving high coverage of the true item. 

To achieve this, we introduce two CP-based objectives:
\begin{itemize}[leftmargin=*]
    \item \textbf{CPS:} Encourages the model to make confident predictions by \emph{minimizing} the average size of the prediction sets.
    \item \textbf{CPD:} Improves the model’s alignment between the predicted set and the true next item, by ``pulling’’ the top-$K$ items closer to the ground truth embedding space.
\end{itemize}
Together, these objectives enable the model to generate narrow, high-confidence recommendation sets, improving user satisfaction by presenting fewer irrelevant items while retaining robust coverage guarantees.

\subsection{Data Splitting}
\label{subsec:data_splitting}
A key aspect of our framework is that it requires \emph{no additional labeled data}; instead, it reuses existing validation data for semi-supervised fine-tuning. While validation sets in recommender systems are traditionally used for early stopping or hyperparameter tuning, our approach leverages them more extensively by partitioning each user’s interaction sequence into two subsets:

\paragraph{Training Sequence.} 
Given a user $u_i$ with an interaction history 
\[
\mathcal{S}^{(i)} = \bigl[v^{(i)}_1, v^{(i)}_2, \dots, v^{(i)}_{T_i}\bigr],
\]
we define the \textbf{training sequence} as
\[
\mathcal{S}^{(i)}_{\text{train}}
= \bigl[v^{(i)}_1, \dots, v^{(i)}_{T_i-2}\bigr].
\]
This truncated sequence is used in the \emph{standard} way to compute supervised next-item prediction losses, such as Cross-Entropy (CE). It allows the model to learn the sequential patterns of user preferences.

\paragraph{Calibration Sequence.}
The \textbf{calibration sequence} is defined as:
\[
\mathcal{S}^{(i)}_{\text{calib}}
= \bigl[v^{(i)}_1, \dots, v^{(i)}_{T_i-1}\bigr].
\]
This sequence extends the training sequence by including one additional interaction, but it does \emph{not} directly use the ground-truth item $v^{(i)}_{T_i}$ in supervised learning. Instead, $\mathcal{S}^{(i)}_{\text{calib}}$ is employed to \emph{compute conformal prediction sets} and evaluate how well the model’s confidence aligns with observed user behavior.

By jointly learning from $\mathcal{S}^{(i)}_{\text{train}}$ (via supervised losses) and $\mathcal{S}^{(i)}_{\text{calib}}$ (via CP-based losses), our framework captures a richer representation of user preferences. This dual approach enables better generalization, reduces overfitting, and aligns the model’s uncertainty with practical user interactions.

\subsection{Conformal Prediction for SRecsys}
\label{subsec:cp_srecsys}
Conformal Prediction (CP) is a statistical framework for constructing \emph{prediction sets} that, with high probability, contain the correct outcome. We briefly recap how CP constructs these sets and explain how we adapt it for sequential recommendation systems (SRecsys). Let $\hat{f}$ denote our sequential recommendation model, which outputs an embedding $\mathbf{H}^{(i)}_{\text{train}}$ (or $\mathbf{H}^{(i)}_{\text{calib}}$) for a user’s interaction sequence. The \emph{nonconformity score} $s(\cdot)$ quantifies how unlikely a candidate item $v_j$ is, based on the model’s understanding of likely next items.

\paragraph{Nonconformity Score.}
The \emph{confidence} in predicting item $v_j$ from a sequence $\mathcal{S}^{(i)}$ is defined as:
\[
\text{conf}\bigl(\mathbf{H}^{(i)}, v_j\bigr) 
= \frac{\exp\!\bigl(r(u_i, v_j)\bigr)}
       {\sum_{v_k \in \mathcal{V}} \exp\!\bigl(r(u_i, v_k)\bigr)},
\]
where $r(u_i, v_j) = \mathbf{H}^{(i)} \cdot \mathbf{e}_{v_j}^\top$ is the relevance score between the sequence embedding $\mathbf{H}^{(i)}$ and the candidate item embedding $\mathbf{e}_{v_j}$. Using this confidence, we define the nonconformity score as:
\[
s\bigl(\mathbf{H}^{(i)}, v_j\bigr) 
= 1 \;-\; \text{conf}\bigl(\mathbf{H}^{(i)}, v_j\bigr).
\]
Intuitively, $s(\mathbf{H}^{(i)}, v_j)$ is large when $v_j$ has a low predicted probability, indicating that the model considers $v_j$ an unlikely next item.

\paragraph{Calibration and Set Construction.}
We adopt Split CP~\cite{papadopoulos2002inductive} to calibrate the model and construct prediction sets:
\begin{enumerate}[leftmargin=*]
    \item \emph{Train:} Train the model $\hat{f}$ on the training sequence $\mathcal{S}_{\text{train}}$ using standard supervised loss functions (e.g., Cross-Entropy).
    \item \emph{Calibrate:} Compute nonconformity scores on the calibration sequence $\mathcal{S}_{\text{calib}}$ to form an empirical distribution of scores for known next items. From this distribution, calculate the $(1-\alpha)$-quantile $\hat{q}_{1-\alpha}$ as a threshold.
    \item \emph{Construct:} Generate a prediction set $\hat{C}_\alpha(\mathcal{S}^{(i)}_{\text{calib}})$ for user $u_i$ by including all items $v_j$ whose nonconformity score satisfies:
    \[
    s\bigl(\mathbf{H}^{(i)}_{\text{calib}}, v_j\bigr) 
    \leq \hat{q}_{1-\alpha}.
    \]
\end{enumerate}
Under mild assumptions of exchangeability, CP guarantees that the true next item $v^{(i)}_{T_i}$ lies in the prediction set $\hat{C}_\alpha(\mathcal{S}^{(i)}_{\text{calib}})$ with a probability of at least $(1-\alpha)$.

\paragraph{Why CP is Useful for SRecsys.}
In contrast to typical ranking approaches that produce a sorted list of items without quantifying uncertainty, CP generates a ``confidence-controlled’’ \emph{subset} of items with a statistical coverage guarantee. This is particularly valuable in user-facing systems, where communicating confidence can enhance trust and engagement. For example, a smaller but confidence-calibrated recommendation set reduces cognitive load for users and fosters reliability.

Our approach leverages CP not just at inference but during training, transforming this coverage notion into a \emph{trainable loss function}. By doing so, the model learns to consistently generate smaller, high-confidence recommendation sets that maintain robust coverage. This integration aligns the model's uncertainty estimation with practical deployment goals, ultimately improving both prediction accuracy and user satisfaction.

\subsection{Conformal Prediction-Based Losses}
\label{subsec:cp_based_losses}
Although Conformal Prediction (CP) is traditionally applied \emph{only} at inference time, we extend its concept of \emph{set construction} to guide the model’s parameter updates during training. Specifically, we introduce two new losses: \textbf{Conformal Prediction Set Size (CPS)} and \textbf{Conformal Prediction Set Distance (CPD)}. These losses not only serve as training objectives to optimize the model’s accuracy and confidence but also act as intrinsic confidence metrics. From a CP perspective, the CPS loss quantifies confidence by penalizing large prediction sets, with smaller sets indicating greater confidence in the model’s predictions. Similarly, the CPD loss measures the alignment of top-ranked items with the ground truth in embedding space, reflecting the reliability of the model’s confidence in its top recommendations. Both losses operate \emph{semi-supervisedly}, leveraging the calibration sequence without requiring explicit ground-truth labels, enabling the model to learn confidence signals directly during training.

\subsubsection{Conformal Prediction Set Size (CPS)}
\label{subsubsec:cps}
\textbf{Motivation.} A fundamental metric in conformal prediction is the \emph{size of the prediction set}, also called \emph{inefficiency}. In recommendation terms, a larger set implies lower model confidence (the model needs to hedge its bets by including many possible items), whereas a smaller set indicates that the model is highly confident in a more selective subset.

\paragraph{Definition.} For a mini-batch of users $B$, each user $u_i$ yields a set $\hat{C}_{\alpha}(\mathcal{S}^{(i)}_{\text{calib}})$. The CPS loss is:
\begin{equation}
\label{eq:cps}
\mathcal{L}_{\text{CPS}}
= \frac{1}{|B|}\sum_{u_i \in B} 
\Bigl|\hat{C}_{\alpha}\bigl(\mathcal{S}^{(i)}_{\text{calib}}\bigr)\Bigr|.
\end{equation}
Minimizing $\mathcal{L}_{\text{CPS}}$ pushes the model to reduce the average set size—i.e., to be confident enough that it only needs to include relatively few items to maintain the same coverage level $(1-\alpha)$.

\paragraph{Example.} 
Suppose for a particular user, the model is highly uncertain and yields a conformal set of size 100. This might be acceptable from a coverage standpoint, but unwieldy for practical recommendations. By training with CPS, the model is incentivized to adjust embedding spaces and item likelihoods so that, say, only 10 items suffice to capture the same coverage. This directly addresses user satisfaction: seeing fewer—but more relevant—items is typically preferable.

\subsubsection{Conformal Prediction Set Distance (CPD)}
\label{subsubsec:cpd}
\textbf{Motivation.} Simply shrinking the set size (via CPS) may inadvertently exclude the actual next item, especially in a large item space. The CPD loss mitigates this risk by ``pulling’’ the most plausible subset of items in $\hat{C}_{\alpha}(\mathcal{S}^{(i)}_{\text{calib}})$ closer to the (unknown) ground truth embedding. Hence, if the model \emph{nearly} captures the correct item within its top candidates, CPD further aligns those candidates more tightly.

\paragraph{Definition.} Denote by
\[
K_i
= \mathrm{TopKClosest}\!\Bigl(
\hat{C}_{\alpha}\bigl(\mathcal{S}^{(i)}_{\text{calib}}\bigr),\, v^{(i)}_{T_i}
\Bigr)
\]
the top-$K$ items (in $\hat{C}_{\alpha}(\mathcal{S}^{(i)}_{\text{calib}})$) closest to the ground-truth item $v^{(i)}_{T_i}$ in terms of cosine similarity. The CPD loss is:
\begin{equation}
\label{eq:cpd}
\mathcal{L}_{\text{CPD}}
= \frac{1}{K \cdot |B|}
\sum_{u_i \in B} 
\sum_{v_k \in K_i} 
\text{CosSim}\bigl(\mathbf{e}_{v_k}, \mathbf{e}_{\,v^{(i)}_{T_i}}\bigr),
\end{equation}
where $\text{CosSim}(\cdot,\cdot)$ is a distance measure in embedding space. Minimizing $\mathcal{L}_{\text{CPD}}$ encourages these $K$ nearest candidates to have smaller distance to $v^{(i)}_{T_i}$, thereby making it more likely that the ground truth remains in or near the conformal set.

\paragraph{Example.}
Suppose the prediction set $\hat{C}_\alpha$ includes two highly probable items that are almost correct but misaligned in embedding space. Over successive training iterations, $\mathcal{L}_{\text{CPD}}$ refines these embeddings, pulling them closer to the ground truth. This reduces the risk of excluding the correct item during set size minimization and ensures higher quality rankings within the set.

\subsection{Overall Fine-Tuning Objective}
\label{subsubsec:overall_loss}
In the fine-tuning stage, we combine the standard supervised loss (e.g., Cross-Entropy, denoted $\mathcal{L}_{\text{pre}}$) with our two conformal losses:
\begin{equation}
\label{eq:cpft}
\mathcal{L}_{\text{CPFT}}
= \mathcal{L}_{\text{pre}}
\;+\; \beta\,\mathcal{L}_{\text{CPS}}
\;+\; \gamma\,\mathcal{L}_{\text{CPD}},
\end{equation}
where $\beta$ and $\gamma$ control the relative influence of CPS and CPD, respectively. By minimizing $\mathcal{L}_{\text{CPFT}}$, the model is explicitly encouraged to:
\begin{itemize}[leftmargin=*]
    \item Maintain high accuracy on the next-item prediction task (\emph{through} $\mathcal{L}_{\text{pre}}$).
    \item Produce narrower, more confident recommendation sets that still retain the correct item with high coverage (\emph{through} $\mathcal{L}_{\text{CPS}}$).
    \item Keep the conformal set items close to the ground truth embedding, thereby reducing the likelihood of excluding the correct item and improving ranking quality (\emph{through} $\mathcal{L}_{\text{CPD}}$).
\end{itemize}

\paragraph{Practical Benefits.}
From a practical perspective, the inclusion of CP-based losses leads to:
\begin{enumerate}[leftmargin=*]
    \item \textbf{Improved User Trust:} Smaller, confidence-calibrated recommendation sets promote a sense of reliability.
    \item \textbf{Reduced Cognitive Load:} Users do not have to sift through extensive candidate lists; the system learns to present fewer, more relevant items.
    \item \textbf{Statistical Guarantees:} If desired, we can still interpret the final predictions through the lens of CP’s coverage guarantees, which are now partially enforced \emph{within} the model’s parameters.
\end{enumerate}
By training \emph{with} CP rather than applying it post-hoc at inference, we align the model’s learning objectives more closely with practical deployment scenarios, ultimately improving both recommendation accuracy and user satisfaction.

\section{Experiments}
In this section, we conduct a comprehensive empirical evaluation of our proposed conformal prediction-based fine-tuning losses across five real-world datasets using four distinct models. 

\subsection{Experimental Setup}
\subsubsection{Dataset}
To rigorously evaluate the performance of our proposed methodology, we conducted experiments on five datasets with four sequential models. Key statistics of the preprocessed datasets are summarized in Tab.~\ref{tab:dataset}. Specifically, we use five publicly available real-world datasets from the Amazon Review Dataset\footnote{https://jmcauley.ucsd.edu/data/amazon/}: Scientific, Pantry, Instruments, and Arts and Office. These datasets have been widely used in previous recommendations system studies~\cite{UnisRec, VQ_Rec, cw:pre-training-graph-neural-network}.

\subsubsection{Evaluation Protocol} 
We adopt the widely accepted leave-one-out (LOO) evaluation methodology to assess the effectiveness of our fine-tuning losses. In this approach, the most recent interaction of a user is used as the test item, while the penultimate interaction serves as the validation item. The model’s performance is evaluated based on its ability to rank the test item among a collection of negative items—items the user has not interacted with. Following established methodologies like SASRec~\cite{sasrec} and UnisRec~\cite{UnisRec}, we employ a full-ranking technique, assessing the model's ability to rank all unobserved items for each user.

As our primary interest lies in top-N item recommendation, we use normalized discounted cumulative gain (NDCG@10,50) and Hit-Rate (HR@10,50) as key evaluation metrics. Additionally, we highlight that the CPFT loss can itself serve as a confidence metric during evaluation. Specifically, the CPFT loss encapsulates the model’s ability to generate compact, confidence-calibrated prediction sets, with smaller losses reflecting higher confidence in the recommendations. By interpreting CPFT loss as a confidence signal, we provide an additional perspective on the model's performance, enabling a more nuanced evaluation of its recommendation quality.

\subsubsection{Baselines} 
In our comprehensive evaluation of conformal prediction-based fine-tuning for sequential recommendation systems, we compare our approach against a diverse range of models to establish its effectiveness. Among these, SASRec~\cite{sasrec}, $\text{S}^3$-Rec~\cite{s3rec}, and FDSA~\cite{fdsa} are categorized as single-domain models. SASRec employs directional self-attention to capture item correlations within sequences effectively. $\text{S}^3$-Rec, on the other hand, utilizes a bidirectional Transformer encoder during its pretraining stage, enabling it to capture multi-view correlations. FDSA enhances item representation by concatenating item features with item ID embeddings. To further analyze the model-agnostic capabilities of our fine-tuning losses, we include UnisRec~\cite{UnisRec} in our set of baseline models. UnisRec stands out as a pre-trained universal sequence representation model, known for its efficiency in transferring to new recommendation domains or platforms with minimal parameter adjustments. In summary, our baseline models encompass a broad spectrum: SASRec, which is an ID-based single-domain model; $\text{S}^3$-Rec, a single-domain model that integrates pre-training and fine-tuning; FDSA, which combines item features and ID information; and UnisRec, a versatile model for universal sequence representation. This selection enables a thorough examination of the performance of our fine-tuning losses across varied model architectures and recommendation scenarios.

\subsubsection{Hyper-parameters and Grid Search} To guarantee fairness and ensure the reliability of our experimental results, we integrated our proposed framework with RecBole~\cite{recbole[1.0]}. In the case of SASRec, $\text{S}^3$Rec, and FDSA, we implemented on a comprehensive hyper-parameter optimization process. This involved experimenting with various learning rates (0.001, 0.0005, 0.0001), setting the embedding dimension to 300, and testing different configurations for the number of layers (1, 2, 4) and heads (1, 2, 4). We standardized the batch size across these models at 512. For UnisRec, we adhered to the default settings as prescribed by RecBole. Additionally, we utilized the pre-trained checkpoints provided by the original authors of UnisRec to ensure consistency and leverage the model's full potential.

\begin{table}[]
\centering
\caption{The Statistics of Preprocessed Datasets: $\textbf{'Avg.U'}$ represents the average number of interactions per user, $\textbf{'Avg.I'}$ signifies the average number of interactions per item.}
\label{tab:dataset}
\resizebox{\linewidth}{!}{%
\begin{tabular}{llllll}
\toprule
 & \textbf{Scientific} & \textbf{Pantry} & \textbf{Instruments} & \textbf{Arts} & \textbf{Office} \\ \hline
\textbf{\#Users} & 8,443 & 13,102 & 24,963 & 45,487 & 87,347\\
\textbf{\#Items} & 4,386 & 4,899 & 9,965 & 21,020 & 25,987\\
\textbf{\#Inters} & 50,985 & 113,861 & 183,964 & 349,664 & 597,491 \\
\textbf{\#Avg.U} & 6.039 & 8.691 & 7.370 & 7.687 & 6.841\\
\textbf{\#Avg.I} & 11.683 & 23.246 & 18.483 & 16.643 & 23.080\\
\bottomrule
\end{tabular}%
}
\end{table}

\subsection{Overall Performance}

\begin{table*}[]
\centering
\caption{Performance Comparison of SRecsys: Standard Training vs. CPFT. This table contrasts the performance of various sequential recommendation models, where 'SR' denotes models trained via standard methods, and 'SR$_{\text{CPFT}}$' signifies models fine-tuned with our proposed loss combined with Cross-Entropy (CE) loss. The highest performance results are highlighted in bold. Improvements exceeding 7\%, 5\%, and 3\% are denoted in {\color[HTML]{CB0000}red}, {\color[HTML]{F56B00}orange}, and {\color[HTML]{3531FF}blue}, respectively. Total$_{\text{Avg-Improv}}$=4.551\%, R@10$_{\text{Avg-Improv}}$=4.994\%,N@10$_{\text{Avg-Improv}}$=-0.004\%,R@50$_{\text{Avg-Improv}}$=6.216\% ,and N@50$_{\text{Avg-Improv}}$=4.630\%}
\label{tab:experiment}
\resizebox{\textwidth}{!}{%
\begin{tabular}{|l|llll|llll|llll|llll|llll|}
\hline
\multicolumn{1}{|c|}{} & \multicolumn{4}{c|}{\textbf{Scientific}} & \multicolumn{4}{c|}{\textbf{Pantry}} & \multicolumn{4}{c|}{\textbf{Instruments}} & \multicolumn{4}{c|}{\textbf{Arts}} & \multicolumn{4}{c|}{\textbf{Office}} \\ \hline
\multicolumn{1}{|c|}{} & \multicolumn{1}{l|}{R@10} & \multicolumn{1}{l|}{N@10} & \multicolumn{1}{l|}{R@50} & N@50 & \multicolumn{1}{l|}{R@10} & \multicolumn{1}{l|}{N@10} & \multicolumn{1}{l|}{R@50} & N@50 & \multicolumn{1}{l|}{R@10} & \multicolumn{1}{l|}{N@10} & \multicolumn{1}{l|}{R@50} & N@50 & \multicolumn{1}{l|}{R@10} & \multicolumn{1}{l|}{N@10} & \multicolumn{1}{l|}{R@50} & N@50 & \multicolumn{1}{l|}{R@10} & \multicolumn{1}{l|}{N@10} & \multicolumn{1}{l|}{R@50} & N@50 \\ \hline
SASRec & \multicolumn{1}{l|}{0.1073} & \multicolumn{1}{l|}{0.0555} & \multicolumn{1}{l|}{0.2078} & 0.0767 & \multicolumn{1}{l|}{0.0511} & \multicolumn{1}{l|}{0.0226} & \multicolumn{1}{l|}{0.1391} & 0.0416 & \multicolumn{1}{l|}{0.1127} & \multicolumn{1}{l|}{0.0697} & \multicolumn{1}{l|}{0.2024} & 0.0891 & \multicolumn{1}{l|}{0.1066} & \multicolumn{1}{l|}{0.0595} & \multicolumn{1}{l|}{0.1978} & 0.0794 & \multicolumn{1}{l|}{0.1159} & \multicolumn{1}{l|}{0.0783} & \multicolumn{1}{l|}{0.1763} & 0.0915 \\
SASRec$_{\text{CPFT}}$ & \multicolumn{1}{l|}{{\color[HTML]{F56B00} \textbf{0.1138}}} & \multicolumn{1}{l|}{{\color[HTML]{3531FF} \textbf{0.0577}}} & \multicolumn{1}{l|}{{\color[HTML]{CB0000} \textbf{0.2228}}} & {\color[HTML]{F56B00} \textbf{0.0813}} & \multicolumn{1}{l|}{{\color[HTML]{F56B00} \textbf{0.0543}}} & \multicolumn{1}{l|}{{\color[HTML]{3531FF} \textbf{0.0237}}} & \multicolumn{1}{l|}{{\color[HTML]{CB0000} \textbf{0.1494}}} & {\color[HTML]{F56B00} \textbf{0.0441}} & \multicolumn{1}{l|}{{\color[HTML]{F56B00} \textbf{0.1188}}} & \multicolumn{1}{l|}{{\color[HTML]{000000} \textbf{0.0703}}} & \multicolumn{1}{l|}{{\color[HTML]{CB0000} \textbf{0.2183}}} & {\color[HTML]{3531FF} \textbf{0.0918}} & \multicolumn{1}{l|}{{\color[HTML]{F56B00} \textbf{0.1122}}} & \multicolumn{1}{l|}{{\color[HTML]{F56B00} \textbf{0.0630}}} & \multicolumn{1}{l|}{{\color[HTML]{3531FF} \textbf{0.2073}}} & {\color[HTML]{F56B00} \textbf{0.0836}} & \multicolumn{1}{l|}{\textbf{0.1183}} & \multicolumn{1}{l|}{\textbf{0.0806}} & \multicolumn{1}{l|}{{\color[HTML]{F56B00} \textbf{0.1816}}} & {\color[HTML]{3531FF} \textbf{0.0944}} \\ \hline
$\text{S}^3$-Rec & \multicolumn{1}{l|}{0.0695} & \multicolumn{1}{l|}{0.0373} & \multicolumn{1}{l|}{0.1597} & 0.0567 & \multicolumn{1}{l|}{0.0410} & \multicolumn{1}{l|}{0.0196} & \multicolumn{1}{l|}{0.1357} & 0.0399 & \multicolumn{1}{l|}{0.0984} & \multicolumn{1}{l|}{0.0576} & \multicolumn{1}{l|}{0.1893} & 0.0754 & \multicolumn{1}{l|}{0.0830} & \multicolumn{1}{l|}{\textbf{0.0553}} & \multicolumn{1}{l|}{0.1593} & \textbf{0.0720} & \multicolumn{1}{l|}{0.1036} & \multicolumn{1}{l|}{\textbf{0.0750}} & \multicolumn{1}{l|}{0.1601} & \textbf{0.0872} \\
$\text{S}^3$-Rec$_{\text{CPFT}}$ & \multicolumn{1}{l|}{\textbf{0.0711}} & \multicolumn{1}{l|}{{\color[HTML]{3531FF} \textbf{0.0385}}} & \multicolumn{1}{l|}{\textbf{0.1623}} & \textbf{0.0581} & \multicolumn{1}{l|}{{\color[HTML]{F56B00} \textbf{0.0431}}} & \multicolumn{1}{l|}{{\color[HTML]{3531FF} \textbf{0.0204}}} & \multicolumn{1}{l|}{\textbf{0.1376}} & {\color[HTML]{CB0000} \textbf{0.0431}} & \multicolumn{1}{l|}{{\color[HTML]{CB0000} \textbf{0.1055}}} & \multicolumn{1}{l|}{{\color[HTML]{3531FF} \textbf{0.0596}}} & \multicolumn{1}{l|}{\textbf{0.1938}} & {\color[HTML]{3531FF} \textbf{0.0787}} & \multicolumn{1}{l|}{{\color[HTML]{CB0000} \textbf{0.0941}}} & \multicolumn{1}{l|}{0.0526} & \multicolumn{1}{l|}{{\color[HTML]{CB0000} \textbf{0.1827}}} & 0.0719 & \multicolumn{1}{l|}{\textbf{0.1066}} & \multicolumn{1}{l|}{0.0733} & \multicolumn{1}{l|}{{\color[HTML]{3531FF} \textbf{0.1652}}} & 0.0861 \\ \hline
FDSA & \multicolumn{1}{l|}{0.0879} & \multicolumn{1}{l|}{0.0592} & \multicolumn{1}{l|}{0.1709} & 0.0769 & \multicolumn{1}{l|}{0.0388} & \multicolumn{1}{l|}{0.0216} & \multicolumn{1}{l|}{0.1070} & 0.0362 & \multicolumn{1}{l|}{0.1080} & \multicolumn{1}{l|}{0.0812} & \multicolumn{1}{l|}{0.1930} & 0.0995 & \multicolumn{1}{l|}{0.1006} & \multicolumn{1}{l|}{0.0725} & \multicolumn{1}{l|}{0.1784} & 0.0894 & \multicolumn{1}{l|}{0.1131} & \multicolumn{1}{l|}{0.1663} & \multicolumn{1}{l|}{0.0882} & 0.0997 \\
FDSA$_{\text{CPFT}}$ & \multicolumn{1}{l|}{{\color[HTML]{CB0000} \textbf{0.0951}}} & \multicolumn{1}{l|}{{\color[HTML]{3531FF} \textbf{0.0611}}} & \multicolumn{1}{l|}{{\color[HTML]{CB0000} \textbf{0.1847}}} & {\color[HTML]{F56B00} \textbf{0.0804}} & \multicolumn{1}{l|}{{\color[HTML]{CB0000} \textbf{0.0441}}} & \multicolumn{1}{l|}{{\color[HTML]{CB0000} \textbf{0.0242}}} & \multicolumn{1}{l|}{{\color[HTML]{CB0000} \textbf{0.1263}}} & {\color[HTML]{CB0000} \textbf{0.0418}} & \multicolumn{1}{l|}{{\color[HTML]{3531FF} \textbf{0.1149}}} & \multicolumn{1}{l|}{{\color[HTML]{3531FF} \textbf{0.0843}}} & \multicolumn{1}{l|}{{\color[HTML]{F56B00} \textbf{0.1999}}} & {\color[HTML]{3531FF} \textbf{0.1027}} & \multicolumn{1}{l|}{{\color[HTML]{3531FF} \textbf{0.1046}}} & \multicolumn{1}{l|}{\textbf{0.0738}} & \multicolumn{1}{l|}{{\color[HTML]{3531FF} \textbf{0.1866}}} & \textbf{0.0916} & \multicolumn{1}{l|}{\textbf{0.1139}} & \multicolumn{1}{l|}{\textbf{0.1697}} & \multicolumn{1}{l|}{\textbf{0.0883}} & \textbf{0.1001} \\ \hline
UnisRec & \multicolumn{1}{l|}{0.1233} & \multicolumn{1}{l|}{0.0639} & \multicolumn{1}{l|}{0.2361} & 0.0898 & \multicolumn{1}{l|}{0.0723} & \multicolumn{1}{l|}{0.0327} & \multicolumn{1}{l|}{0.1883} & 0.0569 & \multicolumn{1}{l|}{0.1258} & \multicolumn{1}{l|}{0.0718} & \multicolumn{1}{l|}{0.2358} & 0.0956 & \multicolumn{1}{l|}{0.1195} & \multicolumn{1}{l|}{0.0662} & \multicolumn{1}{l|}{0.2252} & 0.0892 & \multicolumn{1}{l|}{0.1276} & \multicolumn{1}{l|}{0.0822} & \multicolumn{1}{l|}{0.2000} & 0.0986 \\
UnisRec$_{\text{CPFT}}$ & \multicolumn{1}{l|}{{\color[HTML]{3531FF} \textbf{0.1285}}} & \multicolumn{1}{l|}{{\color[HTML]{CB0000} \textbf{0.0695}}} & \multicolumn{1}{l|}{{\color[HTML]{F56B00} \textbf{0.2522}}} & {\color[HTML]{CB0000} \textbf{0.0966}} & \multicolumn{1}{l|}{{\color[HTML]{3531FF} \textbf{0.0758}}} & \multicolumn{1}{l|}{{\color[HTML]{F56B00} \textbf{0.0345}}} & \multicolumn{1}{l|}{\textbf{0.1916}} & {\color[HTML]{3531FF} \textbf{0.0595}} & \multicolumn{1}{l|}{{\color[HTML]{3531FF} \textbf{0.1314}}} & \multicolumn{1}{l|}{{\color[HTML]{F56B00} \textbf{0.0760}}} & \multicolumn{1}{l|}{{\color[HTML]{3531FF} \textbf{0.2442}}} & {\color[HTML]{F56B00} \textbf{0.1005}} & \multicolumn{1}{l|}{{\color[HTML]{CB0000} \textbf{0.1282}}} & \multicolumn{1}{l|}{{\color[HTML]{CB0000} \textbf{0.0711}}} & \multicolumn{1}{l|}{{\color[HTML]{CB0000} \textbf{0.2413}}} & {\color[HTML]{CB0000} \textbf{0.0958}} & \multicolumn{1}{l|}{\textbf{0.1296}} & \multicolumn{1}{l|}{\textbf{0.0843}} & \multicolumn{1}{l|}{\textbf{0.2019}} & \textbf{0.1001} \\ \hline
\end{tabular}%
}
\end{table*}

We evaluated our proposed method against baseline models in five real-world data sets, integrating CE loss with our proposed CPS and CPD losses for fine-tuning. The comparative results are detailed in Tab.~\ref{tab:experiment}. 
Across all datasets and baseline methods, our fine-tuning framework achieved an average improvement of 4.551\% across all metrics. Notably, the Recall metrics saw the most significant enhancement, aligning with our objective to boost recall by enhancing efficiency while maintaining a certain level of confidence. On the other hand, our CP loss in the form of sets also contributes to the improvement of NDCG metrics. Furthermore, our fine-tuning framework demonstrated improvements across all datasets and models, underscoring its versatility. Each model, with its distinct training approach—such as SASRec's use of self-attention for learning item representations and UnisRec's fixed item representations with adaptable mapping to new spaces. This versatility confirms the model-agnostic nature of our framework.

\subsection{Further Analysis}
\subsubsection{Ablation Study}
\begin{table}[]
\centering
\caption{Ablation Study on CP-Loss Variants: This table compares the SASRec performance of different CP-loss combinations during the fine-tuning phase on the "Scientific" and "Office" datasets. Each square bracket denotes a unique set of fine-tuning loss configurations. The best results for each metric in each dataset are highlighted in bold. }
\label{tab:ablation}
\resizebox{\linewidth}{!}{%
\begin{tabular}{l|llll|llll}
\hline
 & \multicolumn{4}{c|}{\textbf{Scientific}} & \multicolumn{4}{c}{\textbf{Office}} \\ \cline{2-9} 
Loss Config & R@10 & N@10 & R@50 & N@50 & R@10 & N@10 & R@50 & N@50 \\ \hline
{[}CE{]} & 0.1073 & 0.0555 & 0.2078 & 0.0767 & 0.1159 & 0.0783 & 0.1763 & 0.0915 \\
{[}CPS{]} & 0.0309 & 0.0173 & 0.0957 & 0.0309 & 0.0244 & 0.0108 & 0.0657 & 0.0197 \\
{[}CE,CPS{]} & 0.1111 & 0.0575 & 0.2135 & 0.0797 & 0.1174 & 0.0790 & 0.1771 & 0.0921 \\
{[}CPS,CPD{]} & 0.0336 & 0.0185 & 0.1156 & 0.0359 & 0.0272 & 0.0119 & 0.0775 & 0.0229 \\
{[}CE,CPS,CPD{]} & \textbf{0.1138} & \textbf{0.0577} & \textbf{0.2228} & \textbf{0.0813} & \textbf{0.1182} & \textbf{0.0796} & \textbf{0.1830} & \textbf{0.0928} \\ \hline
\end{tabular}%
}
\end{table}

In Table \ref{tab:ablation}, we examine the impact of our proposed losses on overall performance by testing five different fine-tuning loss configurations: (1) [CE] employing only CE loss, (2) [CPS] using solely CPS loss, (3) [CE, CPS] a combination of CE and CPS losses, (4) [CPS, CPD] applying both our proposed losses, and (5) [CE, CPS, CPD] incorporating all three losses for fine-tuning.
Given that CPD builds upon the set generated by CPS and serves as its complement, it is not utilized independently during the fine-tuning phase. The analysis reveals that each proposed component contributes positively to enhancing recommendation performance. However, it's important to note that our losses are not standalone solutions but are designed to work in conjunction with CE loss. While CE directly targets the accuracy of the recommendation, our losses function as regularizers, increasing the confidence in the prediction.

\subsubsection{Validating Confidence Enhancement Effectiveness}
\begin{figure}
    \begin{subfigure}{\linewidth}
        \includegraphics[width=\linewidth]{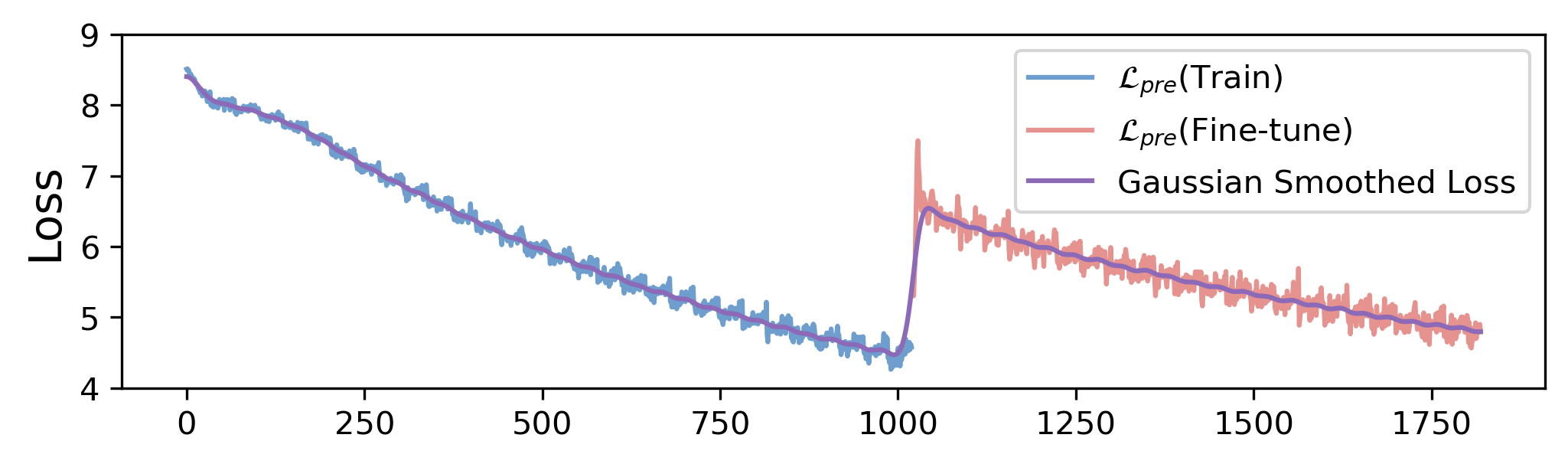}
        \caption{Prediction Loss Across Training and CPFT}
        \label{fig:ce_loss}
    \end{subfigure}
    \hfill
    \begin{subfigure}{\linewidth}
        \includegraphics[width=\linewidth]{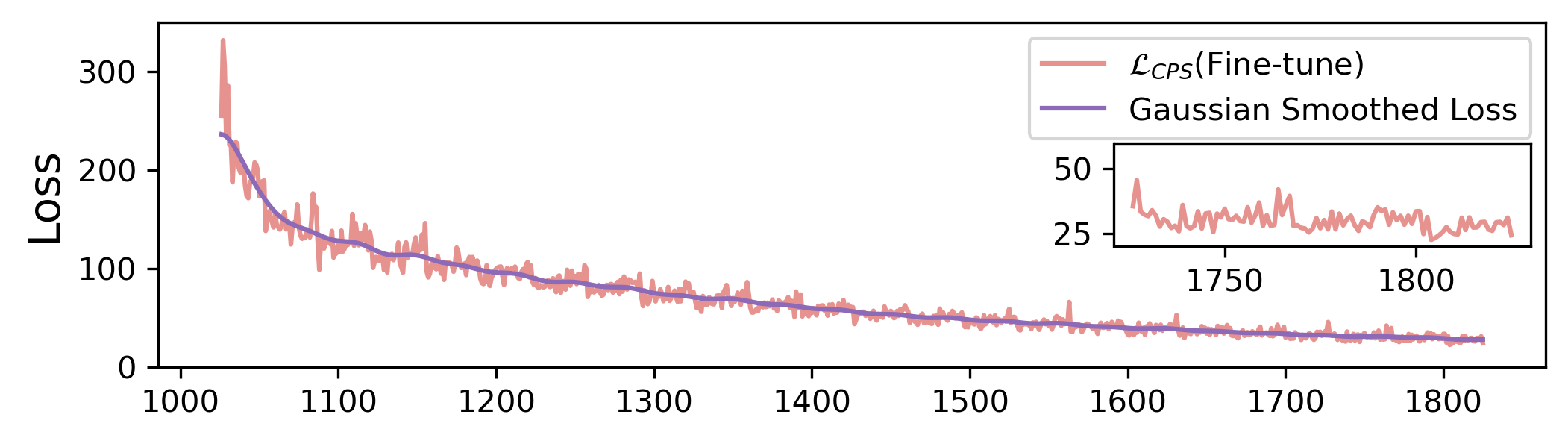}
        \caption{CPS loss during CPFT}
        \label{fig:cps_loss}
    \end{subfigure}
    \begin{subfigure}{\linewidth}
        \includegraphics[width=\linewidth]{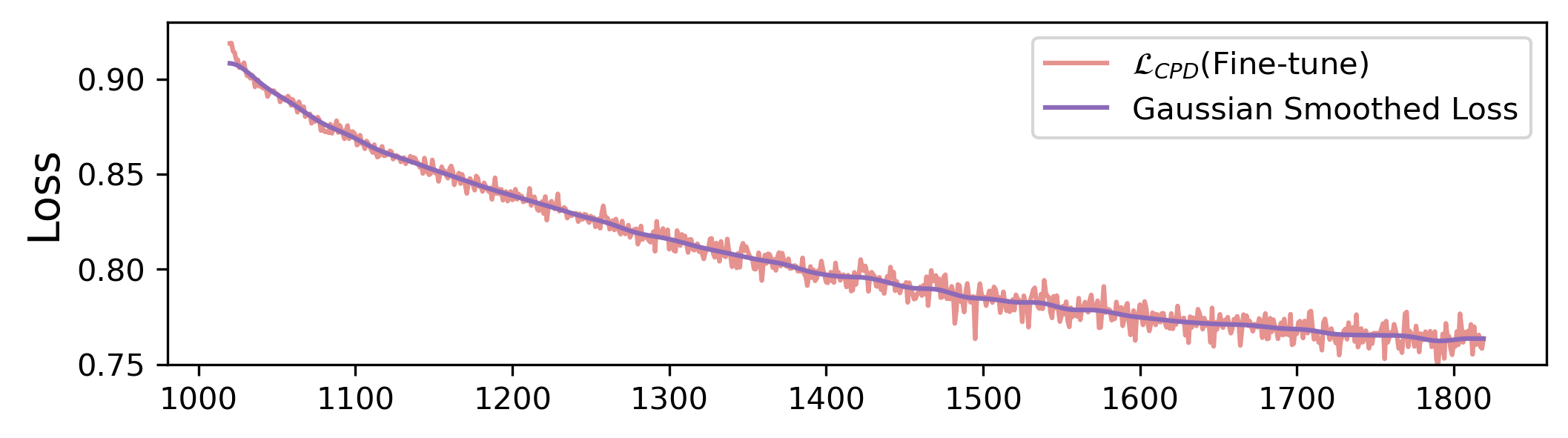}
        \caption{CPD loss during CPFT}
        \label{fig:cpd_loss}
    \end{subfigure}
    \begin{subfigure}{\linewidth}
        \includegraphics[width=\linewidth]{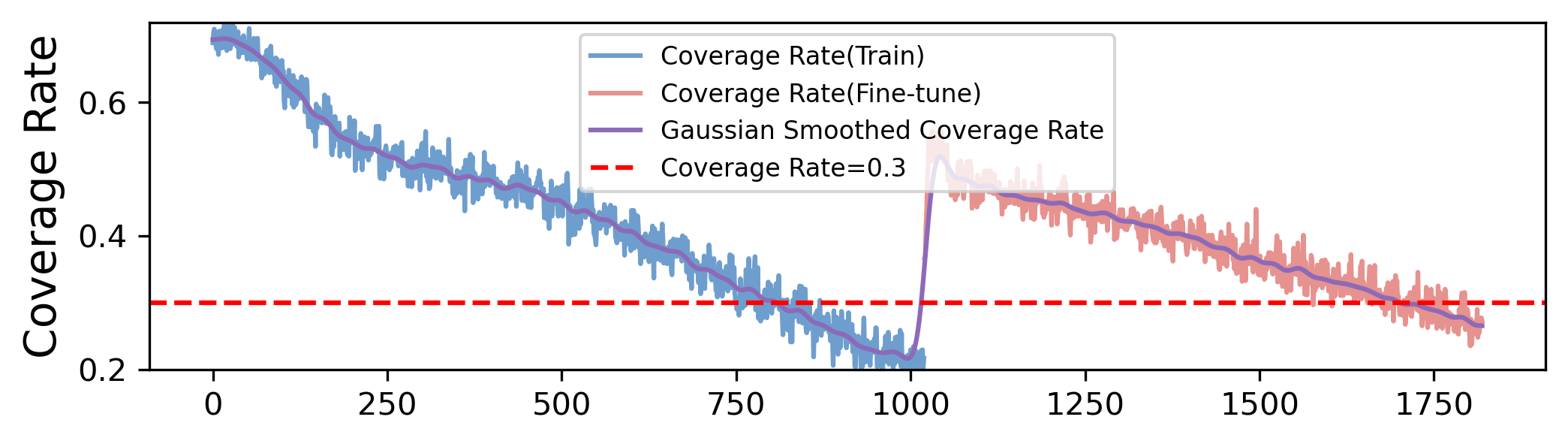}
        \caption{Coverage Rate Across Training and CPFT}
        \label{fig:coverage}
    \end{subfigure}
    \caption{Performance Metrics of the SASRec Model During Training and CPFT. The model parameters were set with a learning rate of 0.0005, $\alpha=0.7$, $\beta=10$, $\gamma=1$, and $\text{TopKClosest}=10$ in the Scientific dataset. Lower values of the CPS loss and CPD loss indicate higher confidence.}
    \label{fig:loss_monitor}
\end{figure}

To assess the efficacy of our framework in boosting model confidence, we tracked metrics throughout the training and fine-tuning phases. Figure~\ref{fig:loss_monitor} illustrates the progression of $\mathcal{L}_{pre}$, $\mathcal{L}_{CPS}$, $\mathcal{L}_{CPD}$, and the coverage rate. Figure~\ref{fig:cps_loss} reveals that the size of the prediction set decreases from more than 300 to approximately 25 during training, indicating a more focused set of predictions. As confidence is normalized, higher confidence levels among the top items prevent other items with lower confidence from exceeding the threshold and being included in the prediction set. Therefore, a reduction in the size of the prediction set is associated with the confidence levels of the top items predicted by the model, supported by our case study in Tab.~\ref{tab:case_study}. Conventionally, only under the same coverage rate, a smaller set size indicates higher confidence~\cite{angelopoulos2021gentle,huang2023graph}. Interestingly, as shown in Figure~\ref{fig:coverage}, along with the increased confidence, the coverage rate decreases and converges at around the predefined level. This is because CPS is co-trained with CE, which aims to lift the confidence in predicting training items with ground-truth labels, reflected by an increased threshold. Therefore, even with increased confidence, the calibration items are compared with a stricter threshold, leading to a decreasing coverage rate. We believe that this adapting threshold is beneficial for the overall training in SRecsys, though the exchangeability is not strictly satisfied.



\subsubsection{Hyper-parameter Sensitivity}

\begin{table}[]
\centering
\caption{Sensitivity Analysis of CPFT Hyper-Parameters: An examination using the SASRec model on the Scientific dataset, with optimal settings (Tab.~\ref{tab:experiment}) emphasized in \textbf{bold}. This analysis highlights the influence of $\alpha$, $\beta$, and $\gamma$ on the performance of the model and the decreasing effect of TopKClosest beyond a threshold. HP stands for hyper-parameters.}
\label{tab:sensitivaty}
\resizebox{\linewidth}{!}{%
\begin{tabular}{|l|cccc|cccc|}
\hline
\multicolumn{1}{|c|}{} & \multicolumn{4}{c|}{Error Rate $\alpha$} & \multicolumn{4}{c|}{TopKClosest} \\ \hline
HP & \multicolumn{1}{c|}{0.1} & \multicolumn{1}{c|}{\textbf{0.3}} & \multicolumn{1}{c|}{0.5} & 0.7 & \multicolumn{1}{c|}{1} & \multicolumn{1}{c|}{\textbf{10}} & \multicolumn{1}{c|}{20} & 50 \\ \hline
R@10 & \multicolumn{1}{c|}{0.1059} & \multicolumn{1}{c|}{\textbf{0.1138}} & \multicolumn{1}{c|}{0.1097} & 0.1104 & \multicolumn{1}{c|}{0.1086} & \multicolumn{1}{c|}{\textbf{0.1138}} & \multicolumn{1}{c|}{0.1128} & 0.1123 \\
N@10 & \multicolumn{1}{c|}{0.0549} & \multicolumn{1}{c|}{\textbf{0.0577}} & \multicolumn{1}{c|}{0.0552} & 0.0554 & \multicolumn{1}{c|}{0.0553} & \multicolumn{1}{c|}{\textbf{0.0577}} & \multicolumn{1}{c|}{0.0572} & 0.0571 \\ \hline
\multicolumn{1}{|c|}{} & \multicolumn{4}{c|}{Size Weight $\beta$} & \multicolumn{4}{c|}{Distance Weight $\gamma$} \\ \hline
HP & \multicolumn{1}{c|}{0.1} & \multicolumn{1}{c|}{1} & \multicolumn{1}{c|}{\textbf{10}} & 20 & \multicolumn{1}{c|}{0.1} & \multicolumn{1}{c|}{\textbf{1}} & \multicolumn{1}{c|}{2} & 5 \\ \hline
R@10 & \multicolumn{1}{c|}{0.1078} & \multicolumn{1}{c|}{0.1098} & \multicolumn{1}{c|}{\textbf{0.1138}} & 0.1104 & \multicolumn{1}{c|}{0.1115} & \multicolumn{1}{c|}{\textbf{0.1138}} & \multicolumn{1}{c|}{0.1122} & 0.1108 \\
N@10 & \multicolumn{1}{c|}{0.0553} & \multicolumn{1}{c|}{0.0568} & \multicolumn{1}{c|}{\textbf{0.0577}} & 0.0557 & \multicolumn{1}{c|}{0.057} & \multicolumn{1}{c|}{\textbf{0.0577}} & \multicolumn{1}{c|}{0.0576} & 0.0574 \\ \hline
\end{tabular}%
}
\end{table}

In Tab.~\ref{tab:sensitivaty}, we explore the sensitivity of individual hyperparameters for CP-based losses using the SASRec model on the Scientific dataset, with key parameters from the main study (Tab.~\ref{tab:experiment}) highlighted in \textbf{bold}. Our findings indicate that $\alpha=0.3$ optimally suits the SRecsys tasks. The impact of TopKClosest diminishes for values beyond 20, aligning with $\mathcal{L}_{CPS}$ which typically narrows the prediction set size to 20-30. For $\mathcal{L}_{CPS}$, the size weight $\beta$ significantly influences model performance, whereas $\gamma$, the hyperparameter for $\mathcal{L}_{CPD}$, shows a minimal effect.

\begin{table*}
\centering
\caption{Comparative Performance Analysis of SASRec and CPFT Models With and Without Validation Data}
\label{tab:valid-exp}
\resizebox{\textwidth}{!}{%
\begin{tabular}{|l|llll|llll|llll|llll|llll|}
\hline
\multicolumn{1}{|c|}{} & \multicolumn{4}{c|}{Scientific} & \multicolumn{4}{c|}{Pantry} & \multicolumn{4}{c|}{Instruments} & \multicolumn{4}{c|}{Arts} & \multicolumn{4}{c|}{Office} \\ \hline
\multicolumn{1}{|c|}{} & \multicolumn{1}{l|}{R@10} & \multicolumn{1}{l|}{N@10} & \multicolumn{1}{l|}{R@50} & N@50 & \multicolumn{1}{l|}{R@10} & \multicolumn{1}{l|}{N@10} & \multicolumn{1}{l|}{R@50} & N@50 & \multicolumn{1}{l|}{R@10} & \multicolumn{1}{l|}{N@10} & \multicolumn{1}{l|}{R@50} & N@50 & \multicolumn{1}{l|}{R@10} & \multicolumn{1}{l|}{N@10} & \multicolumn{1}{l|}{R@50} & N@50 & \multicolumn{1}{l|}{R@10} & \multicolumn{1}{l|}{N@10} & \multicolumn{1}{l|}{R@50} & N@50 \\ \hline
SASRec-$w/o$-valid & \multicolumn{1}{l|}{\textbf{0.1048}} & \multicolumn{1}{l|}{\textbf{0.0536}} & \multicolumn{1}{l|}{\textbf{0.2056}} & \textbf{0.0756} & \multicolumn{1}{l|}{\textbf{0.0499}} & \multicolumn{1}{l|}{\textbf{0.0221}} & \multicolumn{1}{l|}{\textbf{0.1367}} & \textbf{0.0408} & \multicolumn{1}{l|}{\textbf{0.1166}} & \multicolumn{1}{l|}{\textbf{0.0691}} & \multicolumn{1}{l|}{\textbf{0.2124}} & \textbf{0.0898} & \multicolumn{1}{l|}{\textbf{0.1117}} & \multicolumn{1}{l|}{\textbf{0.0621}} & \multicolumn{1}{l|}{\textbf{0.2038}} & \textbf{0.0821} & \multicolumn{1}{l|}{\textbf{0.1177}} & \multicolumn{1}{l|}{\textbf{0.0757}} & \multicolumn{1}{l|}{\textbf{0.1811}} & \textbf{0.0895} \\ 
SASRec-w/th-valid & \multicolumn{1}{l|}{0.1010} & \multicolumn{1}{l|}{0.0526} & \multicolumn{1}{l|}{0.1958} & {0.0733} & \multicolumn{1}{l|}{0.0432} & \multicolumn{1}{l|}{0.0201} & \multicolumn{1}{l|}{0.1281} & {0.0399} & \multicolumn{1}{l|}{0.1153} & \multicolumn{1}{l|}{0.0647} & \multicolumn{1}{l|}{0.2162} & {0.0910} & \multicolumn{1}{l|}{0.1071} & \multicolumn{1}{l|}{0.0607} & \multicolumn{1}{l|}{0.2048} & {0.0814} & \multicolumn{1}{l|}{0.1085} & \multicolumn{1}{l|}{0.0723} & \multicolumn{1}{l|}{0.1756} & {0.0887} \\ \hline
CPFT-w$/$o-valid & \multicolumn{1}{l|}{0.1129} & \multicolumn{1}{l|}{0.0546} & \multicolumn{1}{l|}{0.2153} & \textbf{0.0811} & \multicolumn{1}{l|}{0.0519} & \multicolumn{1}{l|}{0.0221} & \multicolumn{1}{l|}{0.1431} & 0.0412 & \multicolumn{1}{l|}{0.1221} & \multicolumn{1}{l|}{0.0699} & \multicolumn{1}{l|}{0.2224} & \textbf{0.0931} & \multicolumn{1}{l|}{0.1143} & \multicolumn{1}{l|}{0.0653} & \multicolumn{1}{l|}{0.2112} & 0.0845 & \multicolumn{1}{l|}{0.1187} & \multicolumn{1}{l|}{0.0763} & \multicolumn{1}{l|}{\textbf{0.1836}} & 0.0912 \\
CPFT-wth-valid & \multicolumn{1}{l|}{\textbf{0.1138}} & \multicolumn{1}{l|}{\textbf{0.0577}} & \multicolumn{1}{l|}{\textbf{0.2228}} & {0.0813} & \multicolumn{1}{l|}{\textbf{0.0543}} & \multicolumn{1}{l|}{ \textbf{0.0237}} & \multicolumn{1}{l|}{\textbf{0.1494}} & {\textbf{0.0441}} & \multicolumn{1}{l|}{{0.1188}} & \multicolumn{1}{l|}{\textbf{0.0703}} & \multicolumn{1}{l|}{{\textbf{0.2183}}} & {0.0918} & \multicolumn{1}{l|}{\textbf{0.1122}} & \multicolumn{1}{l|}{{ \textbf{0.0630}}} & \multicolumn{1}{l|}{{ \textbf{0.2073}}} & {\textbf{0836}} & \multicolumn{1}{l|}{\textbf{0.1183}} & \multicolumn{1}{l|}{ \textbf{0.0806}} & \multicolumn{1}{l|}{0.1816} & {\textbf{0.0944}} \\ \hline
\end{tabular}%
}
\end{table*}

\subsubsection{Case Study}

\begin{table}[]
\centering
\caption{Top-5 Prediction Set Comparison: Regular CE Training vs. CPFT. The CP fine-tuning method effectively selects four highly relevant items to the ground truth, each with a high confidence score. GT represents the ground truth.}
\label{tab:case_study}
\resizebox{\linewidth}{!}{%
\begin{tabular}{l|llll}
\hline
GT & \multicolumn{4}{c}{Herbal Tea, Raspberry Zinger} \\ \hline
 & \multicolumn{2}{l|}{CE Train, prediction set size=48} & \multicolumn{2}{l}{CP Fine-tune, prediction set size=27} \\ \cline{2-5} 
 Rank & Title & \multicolumn{1}{l|}{Score} & Title & Score \\ \hline
No.1 & Herbal Tea & \multicolumn{1}{l|}{0.0058} & Herbal Tea & 0.0466 \\
No.2 & Purified Water & \multicolumn{1}{l|}{0.0008} & Apple \& Eve 100\% Juice & 0.0346 \\
No.3 & Pop-Tarts Frosted & \multicolumn{1}{l|}{0.0008} & Gatorade Thirst Quencher & 0.0286 \\
No.4 & Hellmann's Mayonnaise & \multicolumn{1}{l|}{0.0007} & San Francisco Bay OneCup & 0.0250 \\
No.5 & Apple \& Eve 100\% Juice & \multicolumn{1}{l|}{0.0006} & Lay's Kettle Cooked Sea Salt & 0.0219 \\ \hline
\end{tabular}%
}
\end{table}
 
In our case study, we analyze the prediction sets by comparing the top-5 ranked items resulting from traditional CE training with those refined by CP-based loss fine-tuning, as detailed in Table~\ref{tab:case_study}. This comparison shows that while CE training and CP fine-tuning successfully place ground truth as the top-ranked item, the confidence scores associated with CP fine-tuning are significantly higher than those from CE training. Additionally, the prediction set generated through CP fine-tuning is almost half the size of the set from CE training, indicating stronger model confidence in its recommendations. Notably, CP fine-tuning adeptly selects items that are more closely related to the ground truth, all of which fall within the beverage category, demonstrating its enhanced relevance-awareness. In contrast, the CE training method tends to include items that are less related and with lower confidence scores. This stark difference underscores the enhanced precision and confidence that CP-based losses fine-tuning introduces, contributing to a more satisfactory user experience.

\section{Conclusion}
In conclusion, our study introduces an innovative approach to enhancing Sequential Recommendation Systems (SRecsys) by integrating Conformal Prediction (CP)-based losses. CPFT, which includes the loss of the conformal prediction set size (CPS) and the loss of the conformal prediction set distance (CPD), is designed to improve the precision and confidence of the recommendations without the need for additional data. By incorporating CP-based losses as regularizers alongside traditional cross-entropy (CE) loss, our framework demonstrates a model-agnostic capability to fine-tune existing recommendation models effectively.

\section{Appendix}
\subsection{Influence of validation data}

We extend the analysis of SASREC and CPFT with and without validation data in Table~\ref{tab:valid-exp}. From the observation,The observed results suggest that incorporating validation data integrating validation data into the fine-tuning process is a nuanced and non-trivial task.modestly enhances CPFT performance without introducing adverse effects, such as overfitting. This pivotal finding opens up the possibility of integrating validation data during the training phase to enhance the performance of recommendation models. To the best of our knowledge, this is the first paper of a Conformal Prediction study that includes validation data during the fine-tuning process.

The observed results indicated a tendency toward overfitting, underscoring that integrating validation data into the fine-tuning process is a nuanced and non-trivial task.

Our experimental results across five real-world datasets and four different SRecsys models validate the efficacy of our approach. The CP-based losses not only contribute to reducing the prediction set size but also ensure that the recommendations are closely aligned with the users' true preferences, thereby increasing the model's confidence in its suggestions.

Building upon our pioneering work in this field, there are numerous exciting avenues for further exploration.
Firstly, extending the application of CP-based losses to various recommendation tasks beyond sequential settings, e.g., graph-based Recsys, may uncover wider applicability and benefits. Secondly, delving into ranking optimization is crucial, as our current approach with set size primarily focuses on recall. Thirdly, substituting with CP adjusted for non-exchangeability \cite{xu2021conformal,barber2023conformal} may bring further performance enhancements. Lastly, investigating the integration of CP-based losses with other prevalent recommendation losses, like BPR~\cite{bpr} and TOP1~\cite{top1}, could offer novel insights and enhancements in recommendation performance.

\subsection{Time Complexity Analysis}
Our proposed CPFT framework introduces conformal prediction (CP)-based computations during the \emph{fine-tuning} stage. Below, we dissect the main components that increase time complexity compared to standard supervised training:

\begin{itemize}[leftmargin=*]
    \item \textbf{Softmax and Comparison Operations:} 
    Computing the softmax confidence scores for each item and comparing them to a threshold requires \(\mathcal{O}(B \times I)\) time per iteration, where \(B\) is the mini-batch size and \(I\) is the number of items.
    
    \item \textbf{Quantile and Calibration Loss Computations:} 
    The quantile calculation used in split conformal prediction scales as \(\mathcal{O}(B \log B)\). Computing the set size (CPS) loss entails counting or summing the conformal set sizes, which scales linearly with both the batch size and the number of items, i.e., \(\mathcal{O}(B \times I)\).
    
    \item \textbf{Distance Loss (CPD) Computation:} 
    The cosine similarity computations for the top-\(K\) items within the conformal set, when done naively, may scale as \(\mathcal{O}(B^2 \times D)\), where \(D\) is the dimensionality of the item embeddings. In practice, this can be reduced through more efficient nearest-neighbor lookups or by leveraging vectorized GPU operations.
\end{itemize}

\paragraph{Practical Mitigation.} 
Two factors alleviate the increased per-epoch complexity:
\begin{enumerate}[leftmargin=*]
    \item \textbf{Vectorized GPU Execution:} Many of these operations (e.g., softmax, comparisons, and cosine similarity) are easily parallelized. Modern deep learning frameworks can significantly accelerate these computations, narrowing the gap between theoretical and actual runtime.
    \item \textbf{Fewer Epochs to Convergence:} In our experiments, CPFT typically converges in fewer epochs compared to training from scratch, since the model is already partially trained. Consequently, the total \emph{wall-clock} time may decrease, even if each epoch is slightly more expensive.
\end{enumerate}

\paragraph{Empirical Results.}
Table~\ref{tab:time_comparison} shows average training times for the largest dataset (Office) across 10 runs. We compare the baseline training from scratch to our CPFT fine-tuning. Notably, CPFT requires only about \(12.27\%\) of the baseline’s training time on average.

\begin{table}[h]
    \centering
    \caption{Training time (in seconds) for baseline vs.\ our CPFT fine-tuning on Office (averaged over 10 runs).}
    \label{tab:time_comparison}
    \begin{tabular}{l|cc}
    \toprule
    \textbf{Model} & \textbf{Train (Baseline)} & \textbf{CPFT} \\
    \midrule
    SASRec  & 7{,}836  & 1{,}106 \\
    FDSA    & 16{,}309 & 2{,}675 \\
    S3Rec   & 56{,}883 & 5{,}542 \\
    UniSRec & 26{,}114 & 2{,}306 \\
    \bottomrule
    \end{tabular}
\end{table}

\vspace{-1em}

\subsection{Why Fine-Tuning Instead of a Single-Stage Integration?}
\label{subsec:why_finetune}
Although it is technically possible to incorporate conformal prediction principles into the \emph{initial} training phase, we opt for a \emph{fine-tuning} design for two key reasons:

\paragraph{1) Original CP is Non-Differentiable.}
Classical conformal prediction~\cite{vovk2005algorithmic} was conceived for inference, offering prediction sets with distribution-free coverage guarantees. Its procedures (quantile computation, set construction) are generally not differentiable, complicating direct integration into end-to-end model training. Moreover, if the model parameters are still highly ``immature'' (e.g., randomly initialized), the conformal sets may be extremely large or yield low-confidence item scores, impeding stable learning.

\paragraph{2) Training-Time Efficiency.}
End-to-end training with CP for large-scale recommendation often proves computationally demanding. Our pilot experiments on smaller datasets show that \emph{incorporating CP from scratch} can yield comparable performance, but at a significantly higher cost when the dataset size grows. By contrast, \emph{fine-tuning} a partially trained model requires only a fraction of the baseline training time, as demonstrated in Table~\ref{tab:time_comparison}. The model already possesses reasonable item embeddings and ranking behavior, so CP-based refinements converge faster and with fewer epochs than a cold start.


    
    

\clearpage

\bibliographystyle{ACM-Reference-Format}
\bibliography{sample-base}

\end{document}